\documentclass[prd,amsmath,amssymb,floatfix,notitlepage]{revtex4}
\usepackage{graphicx}
\usepackage{hyperref}
\usepackage{bm}
\usepackage{amssymb}
\begin{document}
\title{Three Dimensional Lattice Dispersion Relations for Finite Difference Methods in Scalar Field Simulations}

\author{Nikitas Stamatopoulos}
\email{nstamato@dartmouth.edu}
\affiliation{Department of Physics and Astronomy, Dartmouth College,
Hanover, NH 03755, USA}

%\date{December 27, 2011}
\date{\today}
\begin{abstract}
We calculate the lattice dispersion relation for three dimensional simulations of scalar fields. We argue that the mode frequency of scalar fields on the lattice should not be treated as a function of the magnitude of its wavevector but rather of its wavevector decomposition in Fourier space. Furthermore, we calculate how the lattice dispersion relation differs depending on the way that spatial derivatives are discretized when using finite difference methods in configuration space. For applications that require the mode frequency as an average function of the magnitude of the wavevector, we show how to calculate the radially averaged lattice dispersion relation. Finally, we use the publicly available framework LATTICEEASY to show that wrong treatment of dispersion relations in simulations of preheating leads to an inaccurate description of parametric resonance, which results in incorrect calculations of particle number densities during thermalization after inflation.
\end{abstract}

%\keywords{Lattice dispersion relations, Reheating simulations, LATTICEEASY, Inflation}

\maketitle

\section{Introduction}
The application of numerical simulations in theoretical physics has seen a tremendous surge in the last decade. Problems that appeared intractable a few years ago have now been analyzed in depth through advances in hardware design as well as extensive research in numerical methods. Computer programs have been employed to solve problems ranging from amplitude calculations in quantum field theory \cite{smit,numerical_qft} to black hole simulations in numerical general relativity \cite{black_hole_sim}. Particularly when there are nonlinear processes involved, analytical approximations often fail to capture all the feautures of the underlying theories and the use of numerical methods becomes imperative.

The theory of reheating after inflation \cite{basset} has been comprehensively studied using numerical tools which underlined the significance of non-perturbative effects towards thermalization \cite{Felder-equilibrium,preheating}. More recently, considerable interest has been directed towards the study of long-lived, coherent objects called oscillons \cite{oscillons1,oscillons2,oscillons3} and their relevance in cosmological settings \cite{oscillons_cosm1,oscillons_cosm2,oscillons_cosm3} with recent focus in the period of reheating \cite{amin_oscillons,marcelo_oscillons,amin_oscillons2}. In this case, analytical techniques prove to be unsuccessful to account for the spontaneous emergence of oscillons and numerical simulations provide us with the most effective tool to investigate their dynamics. These studies are generally centered around the evolution of scalar fields, but models including vector fields have been employed to examine similar processes in Abelian and non-Abelian Higgs models \cite{joel_higgs,noah_electroweak}. The interest in such simulations of scalar fields has resulted in the release of publicly available numerical frameworks, most notably LATTICEEASY \cite{latticeeasy} and DEFROST \cite{defrost}, as well as the more recent PSpectRe \cite{pspectre}.

In this study, we examine how three dimensional dispersion relations are treated in computer programs that use finite difference methods, including LATTICEEASY and DEFROST. Even though the aforementioned programs deal with discretized space, the dispersion relations employed are not adjusted for finite grid size effects. Meanwhile, they are important in the setup and evolution of the scalar fields, as they control the initial amplitude of fluctuations in the beginning of the simulation, as well as in the definition of the particle number density. As such, we investigate how results of preheating simulations are altered using the correct lattice dispersion relations, focusing on previous work that is based on LATTICEEASY\cite{Felder-equilibrium}. Because LATTICEEASY and DEFROST utilize different methods to discretize the equations of motion, the lattice dispersion relation in each case differs. We show how to correctly calculate them in both approaches. We finally note that even though we focus on preheating simulations, dispersion relations have to be adapted to account for lattice effects in any scalar field simulation using finite different methods \cite{false-vacuum,hindmarsh,shaposhnikov}.

This paper is organized as follows: In Section~\ref{Section:Anisotropic} we derive the lattice dispersion relation for a free massive scalar field using the discretization scheme used in LATTICEEASY. In Section~\ref{Section:Isotropic} we repeat our calculation for a more general, isotropic scheme which is used in DEFROST and compare it to LATTICEEASY. In Section~\ref{Section:Radial} we show how to calculate a radially averaged lattice dispersion relation which is useful whenever mode frequencies are required as a function of the wavevector magnitude. Section~\ref{Section:Preheating} contains the results that we get from simulations of preheating in a $\lambda \phi^4$ chaotic model of inflation using LATTICEEASY with both the continuous and lattice adjusted dispersion relation. We conclude with a summary of our work in Section~\ref{Section:conclusions}.

\section{Lattice Dispersion Relation Using An Anisotropic Discretization Stencil}
\label{Section:Anisotropic}
We consider a free massive scalar field $\phi(\mathbf{x},t)$ of mass $m$ in (3+1)-dimensional flat Minkowski spacetime with Lagrangian density

\begin{equation}
 \mathcal{L}=\frac{1}{2}(\partial_\mu\phi)^2-\frac{1}{2}m^2\phi^2.
\end{equation}
Using $\hbar=c=1$, the equation of motion satisfied by $\phi$ is 

\begin{equation}
\ddot{\phi}-\nabla^2\phi+m^2\phi=0,
\label{EOM}
\end{equation}
where an overdot denotes derivative with respect to time. We uniformly discretize space in a lattice with $N^3$ points, lattice spacing $\Delta x=\Delta y=\Delta z=a$ and employ a second-order accurate expression for the Laplacian to get the discretized equation of motion for the field at site $(i,j,k)$

\begin{equation}
 \ddot{\phi}_{ijk}-\frac{\phi_{(i+1)jk}+\phi_{(i-1)jk}-2\phi_{ijk}}{a^2}-\frac{\phi_{i(j+1)k}+\phi_{i(j-1)k}-2\phi_{ijk}}{a^2}-\frac{\phi_{ij(k+1)}+\phi_{ij(k-1)}-2\phi_{ijk}}{a^2}+m^2\phi_{ijk}=0.
\label{DiscreteAnisotropic}
\end{equation}
We note that this discretization of the Laplacian uses the six nearest neighbors of a point which can lead to anisotropic propagation of errors in the field evolution. This is the discretization scheme used in LATTICEEASY. We seek a solution to Eq.~\ref{DiscreteAnisotropic} of the form 

\begin{equation}
 \phi(\mathbf{x}_{ijk},t)=Ae^{i(\mathbf{k}\cdot{\mathbf{x}_{ijk}}-\omega t)},
\label{SolutionAnisotropic}
\end{equation}
where $\mathbf{x}_{ijk}=(ia,ja,ka)$ and $\mathbf{k}=(k_x,k_y,k_z)$. We impose periodic boundary conditions which restricts the allowed values of $\mathbf{k}$ to $\mathbf{k}=(k_x,k_y,k_z)=\frac{2\pi}{L}(n_x,n_y,n_z)$ where $n_i$ are integers $n_i=-N/2+1\ldots N/2$ and $L=Na$. Plugging Eq.~\ref{SolutionAnisotropic} into Eq.~\ref{DiscreteAnisotropic} and cancelling common factors we get, after some algebra,

\begin{equation}
 \omega^2-\frac{4}{a^2}\left(\sin^2\frac{k_xa}{2}+\sin^2\frac{k_ya}{2}+\sin^2\frac{k_za}{2}\right)-m^2=0,
\end{equation}
which gives the lattice dispersion relation for this discretization scheme

\begin{equation}
 \omega^2(k_x,k_y,k_z)=k_{\textrm{eff}}^2+m^2,
\label{DispersionAnisotropic}
\end{equation}
with
\begin{equation}
 k_{\textrm{eff}}^2= \frac{4}{a^2}\left(\sin^2\frac{k_xa}{2}+\sin^2\frac{k_ya}{2}+\sin^2\frac{k_za}{2}\right).
\label{keffAnis}
\end{equation}
Unlike the continuous dispersion relation $\omega^2=k^2+m^2$, the lattice dispersion relation formulates the mode frequency $\omega$ as a function of the vector decomposition of $\mathbf{k}$ and not the magnitude $k=|\mathbf{k}|$. Two modes $\mathbf{k}_1=(k_x,k_y,k_z)$ and $\mathbf{k}_2=(k_x^*,k_y^*,k_z^*)$ can have $|\mathbf{k}_1|=|\mathbf{k}_2|$ but $\omega(\mathbf{k}_1)\neq\omega(\mathbf{k}_2)$. For small values of $k_i$, Eq.~\ref{DispersionAnisotropic} reduces to the continuous limit which is expected since large wavelength modes are not significantly affected by finite grid size effects. Large values of $k_i$, however, give mode frequencies that are much different to what one would get using $\omega^2=k^2+m^2$.

The validity of Eq.~\ref{DispersionAnisotropic} is demonstrated by solving Eq.~\ref{DiscreteAnisotropic} on a 3d lattice. We use $N=128$, $a=0.5$, $m=1$ and propagate $\phi(\mathbf{x},t)$ in time using a symplectic second-order Velocity-Verlet algorithm. We initialize the field in Fourier space with random amplitudes for each mode and then track the evolution of two modes with the same wavevector magnitude. We have chosen the modes with $\mathbf{k}_1=\frac{2\pi}{L}(40,28,25)$ and $\mathbf{k}_2=\frac{2\pi}{L}(53,14,2)$ for illustration. They both have $|\mathbf{k}_1|=|\mathbf{k}_2|=\frac{2\pi}{L}\sqrt{3009}$. The results are summarized in Table~\ref{TableAnisResults}. Even though both wavevectors have the same magnitude, their numerically calculated frequencies are significantly different. The continuous dispersion relation fails in the numerical analysis, assigning to both modes the same incorrect frequency. The data also shows excellent agreement between Eq.~\ref{DispersionAnisotropic} and the numerically calculated frequencies.

\begin{table}[htdp]
%\begin{center}
%\begin{ruledtabular}
\begin{tabular*}{0.75\textwidth}{@{\extracolsep{\fill}} c|c|c|c} \hline \hline
Wavevector  & $\omega(k_x,k_y,k_z)$& $\omega=\sqrt{k^2+m^2}$  &  Numerical   \\ \hline \hline
$\frac{2\pi}{L}(40,28,25)$   &   $4.8791m$   & $5.4774m$   &   $4.8707m$  \\ \hline
$\frac{2\pi}{L}(53,14,2)$   &   $4.2091m$   & $5.4774m$   &   $4.2169m$   \\ \hline \hline
%C   &   $1/30$   & $1/10$   &   $7/15$ & $-64/15$   \\ \hline \hline
\end{tabular*}
%\end{ruledtabular}
%\end{center}
\caption{Mode frequencies for two modes computed numerically, their values in the continuous case $\omega^2=k^2+m^2$ and when calculated using Eq.~\ref{DispersionAnisotropic}.}
\label{TableAnisResults}
\end{table}%

%\begin{figure}[htbp]
%\includegraphics[scale=0.5]{frequenciesAnis}
%\caption{Numerically computated time evolution for two modes on the lattice. Even though they have the same magnitude, these modes oscillate with different %frequency, given by Eq.~\ref{DiscreteAnisotropic}. We find $\omega(\mathbf{k}_1)=4.8707m$ and $\omega(\mathbf{k}_2)=4.2169m$}
%\label{Fig:FrequenciesAnis}
%\end{figure}

\section{Lattice Dispersion Relation Using Isotropic Discretization}
\label{Section:Isotropic}
\begin{figure}[htbp]
\includegraphics[scale=0.3]{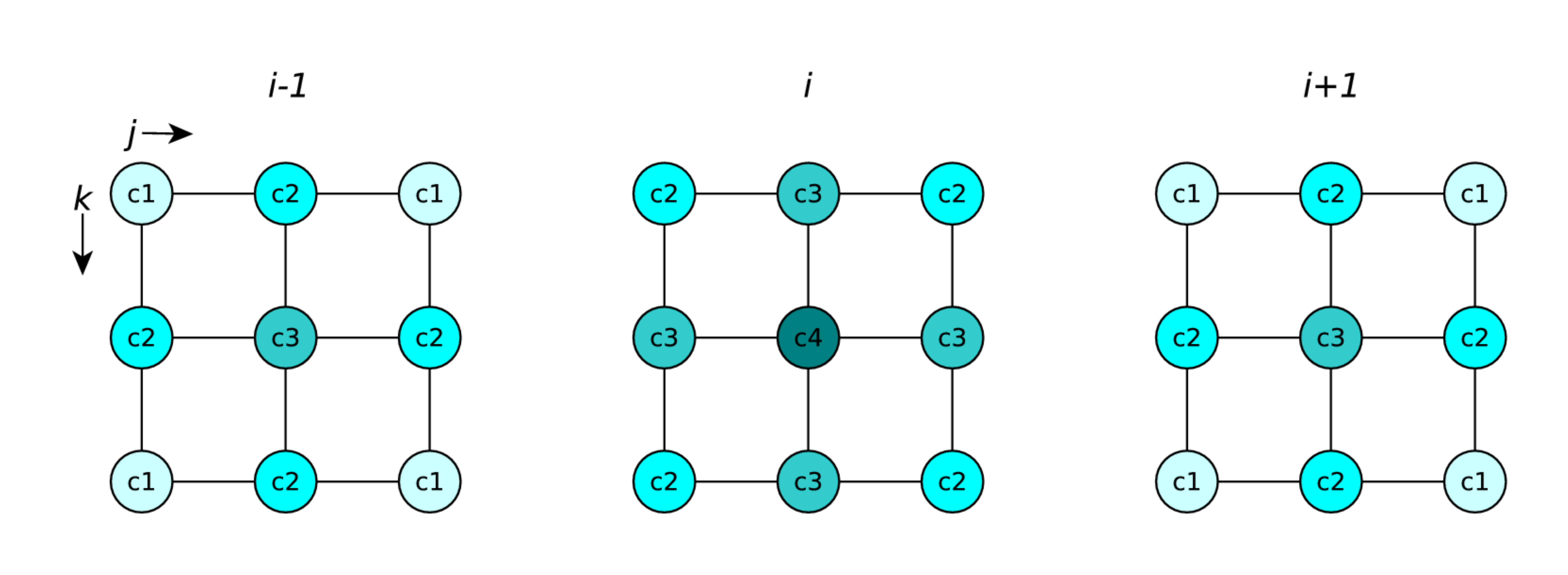}
\caption{The coefficients of the 26 neighbors used in the calculation of the Laplacian at the middle point of the middle two-dimensional slice. Table~\ref{TableIsotropic} shows the values of these coefficients for three isotropic stencils.}
\label{Fig:discrGraph}
 
\end{figure}

The Laplacian term in Eq.~\ref{EOM} can be discretized differently than Eq.~\ref{DiscreteAnisotropic}. Instead of using six neighbors of a lattice point in the discretization of the Laplacian, we can use all 26 neighbors of a $3\times3\times3$ cube around the point. This way, we can derive a family of discretizations which is second-order accurate and fourth-order isotropic \cite{isotropic}. The Laplacian in this case can be written as 

\begin{equation}
 \nabla^2\phi_{ijk}=\frac{D[\phi_{ijk}]}{a^2},
\end{equation}
where $D[\phi_{ijk}]$ is given by

\begin{equation}
 D[\phi_{ijk}]=\sum_{x=i-1}^{i+1}\sum_{y=j-1}^{j+1}\sum_{z=k-1}^{k+1}c_d\phi_{xyz},
\end{equation}
and the coefficients $c_d$ only depend on the distance from $(x,y,z)$ to $(i,j,k)$. The values of these coefficients are displayed in Table~\ref{TableIsotropic} for three isotropic discretization schemes \cite{isotropic}. Fig~\ref{Fig:discrGraph} shows a visualization of the distance coefficients on three two-dimensional slices in the $x$ direction. The standard anisotropic discretization that we employed in the previous section corresponds to $c_1=c_2=0,c_3=1,c_4=-6$. The discrete equation of motion now becomes

\begin{table}[htdp]
%\begin{center}
%\begin{ruledtabular}
\begin{tabular*}{0.5\textwidth}{@{\extracolsep{\fill}} c|c|c|c|c} \hline \hline
coefficient  & $c_1$& $c_2$  &  $c_3$ & $c_4$   \\ \hline \hline
Scheme A   &   $0$   & $1/6$   &   $1/3$ & $-4$  \\ \hline
Scheme B   &   $1/12$   & $0$   &   $2/3$ & $-14/3$  \\ \hline
Scheme C   &   $1/30$   & $1/10$   &   $7/15$ & $-64/15$   \\ \hline \hline
\end{tabular*}
%\end{ruledtabular}
%\end{center}
\caption{Three isotropic discretization schemes for the Laplacian. Scheme C uses all 26 neighbors of a $3\times3\times3$ cube and is used in DEFROST.}
\label{TableIsotropic}
\end{table}%

\begin{equation}
 \ddot{\phi}_{ijk}-\frac{D[\phi_{ijk}]}{a^2}+m^2\phi_{ijk}=0.
\label{DiscreteIsotropic}
\end{equation}
Again, we assume a solution of the form of Eq.~\ref{SolutionAnisotropic} and cancel common terms to get

\begin{equation}
 \omega^2=k_{\textrm{eff}}^2+m^2,
\end{equation}
with $k_{\textrm{eff}}^2$ given by

% \begin{eqnarray}
% k_{\textrm{eff}}^2&=&-\frac{1}{a^2}\left(c_4 +c_3\left[e^{ik_xa}+e^{-ik_xa}+e^{ik_ya}+e^{-ik_ya}+e^{ik_za}+e^{-ik_za}\right]\right.\nonumber\\
%   && \left.+c_2\left[e^{i(k_x+k_y)a}+e^{-i(k_x+k_y)a}+e^{i(k_x+k_z)a}+e^{-i(k_x+k_z)a}+e^{i(k_y+k_z)a}+e^{-i(k_y+k_z)a}\right.\right.\nonumber\\
%   && \left.\left. +e^{i(k_x-k_y)a}+e^{-i(k_x-k_y)a}+e^{i(k_x-k_z)a}+e^{-i(k_x-k_z)a}+e^{i(k_y-k_z)a}+e^{-i(k_y-k_z)a}\right]\right.\nonumber\\
%   && \left.+c_1\left[e^{i(k_x+k_y+k_z)a}+e^{-i(k_x+k_y+k_z)a}+e^{i(k_x+k_y-k_z)a}+e^{-i(k_x+k_y-k_z)a}\right.\right.\nonumber\\
%   && \left.\left.+e^{i(k_x-k_y+k_z)a}+e^{-i(k_x-k_y+k_z)a}+e^{i(-k_x+k_y+k_z)a}+e^{-i(-k_x+k_y+k_z)a}\right]\right).
% \label{keffexponentials}
% \end{eqnarray}
\begin{eqnarray}
k_{\textrm{eff}}^2&=&-\frac{1}{a^2}\bigg(c_4 +c_3\left[e^{ik_xa}+e^{-ik_xa}+e^{ik_ya}+e^{-ik_ya}+e^{ik_za}+e^{-ik_za}\right]\nonumber\\
  && +c_2\left[e^{i(k_x+k_y)a}+e^{-i(k_x+k_y)a}+e^{i(k_x+k_z)a}+e^{-i(k_x+k_z)a}+e^{i(k_y+k_z)a}+e^{-i(k_y+k_z)a}\right.\nonumber\\
  && \left. +e^{i(k_x-k_y)a}+e^{-i(k_x-k_y)a}+e^{i(k_x-k_z)a}+e^{-i(k_x-k_z)a}+e^{i(k_y-k_z)a}+e^{-i(k_y-k_z)a}\right]\nonumber\\
  && +c_1\left[e^{i(k_x+k_y+k_z)a}+e^{-i(k_x+k_y+k_z)a}+e^{i(k_x+k_y-k_z)a}+e^{-i(k_x+k_y-k_z)a}\right.\nonumber\\
  && \left.+e^{i(k_x-k_y+k_z)a}+e^{-i(k_x-k_y+k_z)a}+e^{i(-k_x+k_y+k_z)a}+e^{-i(-k_x+k_y+k_z)a}\right]\bigg).
\label{keffexponentials}
\end{eqnarray}

Using $e^{i\alpha}+e^{-i\alpha}=2\cos \alpha$ and $\cos(\alpha\pm\beta)=\cos\alpha\cos\beta\mp\sin\alpha\sin\beta$, Eq.~\ref{keffexponentials} reduces to

\begin{eqnarray}
 k_{\textrm{eff}}^2&=&-\frac{1}{a^2}(c_4+2c_3[\cos k_xa+\cos k_ya+\cos k_za]+4c_2[\cos k_xa\cos k_ya+\cos k_xa\cos k_za+\cos k_ya\cos k_za]\nonumber\\
&& +8c_1\cos k_xa\cos k_ya \cos k_za)
\end{eqnarray}
and the dispersion relation is then

\begin{eqnarray}
 \omega^2(k_x,k_y,k_z)&=&-\frac{1}{a^2}(c_4+2c_3[\cos k_xa+\cos k_ya+\cos k_za]+4c_2[\cos k_xa\cos k_ya+\cos k_xa\cos k_za+\cos k_ya\cos k_za]\nonumber\\
&& +8c_1\cos k_xa\cos k_ya \cos k_za)+m^2.
\label{DispersionIsotropic}
\end{eqnarray}

\begin{figure}[htbp]
\includegraphics[scale=0.5]{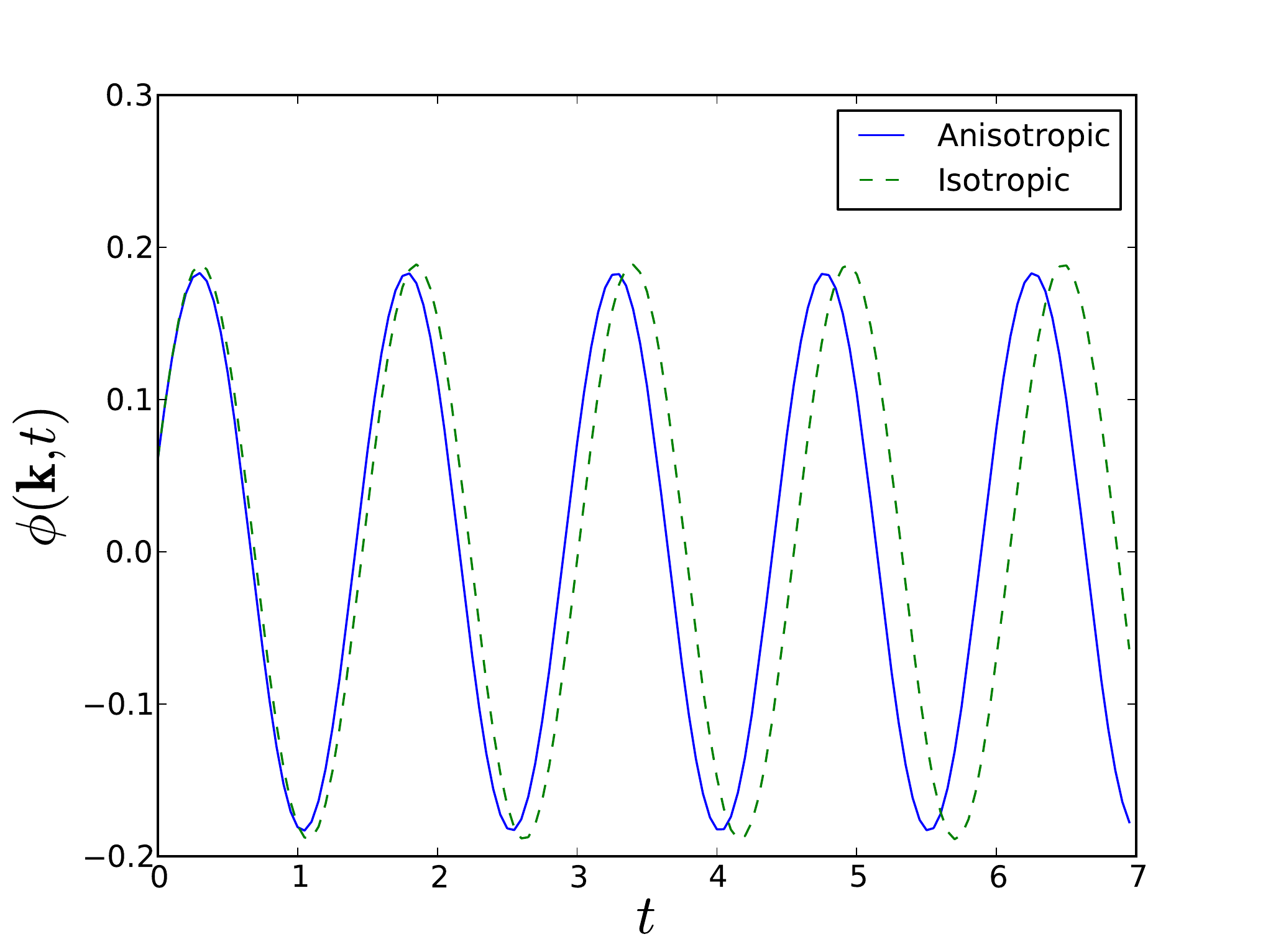}
\caption{Numerical evolution of $\phi(\mathbf{k}_2,t)$ with $\mathbf{k}_2=\frac{2\pi}{L}(53,14,2)$ using anisotropic and isotropic stencils. In the anisotropic case $\omega(\mathbf{k}_2)=4.2169m$ and in the isotropic case $\omega(\mathbf{k}_2)=4.08m$, which illustrates the mode frequency dependence on the discretization stencil used.}
\label{Fig:FrequenciesBoth}
\end{figure}

We perform the same three dimensional simulation as in Section~\ref{Section:Anisotropic} but now using an isotropic discretization of the Laplacian with coefficients $c_1\cdots c_4$ given by the discretization scheme C in Table~\ref{TableIsotropic}. This scheme is used in DEFROST \cite{defrost}. We carry out the same initialization and we focus on the same modes as in Section~\ref{Section:Anisotropic}, $\mathbf{k}_1=\frac{2\pi}{L}(40,28,25)$ and $\mathbf{k}_2=\frac{2\pi}{L}(53,14,2)$. The results are shown in Table~\ref{TableIsResults}. As in Section~\ref{Section:Anisotropic}, the numerically computed mode frequencies are different from the continuous $\omega^2=k^2+m^2$, but agree very well with Eq.~\ref{DispersionIsotropic}. In Fig.~\ref{Fig:FrequenciesBoth} we plot the numerical evolution of $\phi(\mathbf{k}_2,t)$ using the anisotropic discretization stencil of Section~\ref{Section:Anisotropic} and the isotropic discretization stencil C in Table~\ref{TableIsotropic}. It is clear that the dispersion relation is different for these two different discretization schemes.

\begin{table}[htdp]
%\begin{center}
%\begin{ruledtabular}
\begin{tabular*}{0.75\textwidth}{@{\extracolsep{\fill}} c|c|c|c} \hline \hline
Wavevector  & $\omega(k_x,k_y,k_z)$& $\omega=\sqrt{k^2+m^2}$  &  Numerical   \\ \hline \hline
$\frac{2\pi}{L}(40,28,25)$   &   $4.2139m$   & $5.4774m$   &   $4.2084m$  \\ \hline
$\frac{2\pi}{L}(53,14,2)$   &   $4.0703m$   & $5.4774m$   &   $4.08m$   \\ \hline \hline
%C   &   $1/30$   & $1/10$   &   $7/15$ & $-64/15$   \\ \hline \hline
\end{tabular*}
%\end{ruledtabular}
%\end{center}
\caption{Mode frequencies for two modes using the isotropic discretization of Eq.~\ref{DiscreteIsotropic}. We show the numerically computated values, their values in the continuous case $\omega^2=k^2+m^2$ and when calculated using Eq.~\ref{DispersionIsotropic}.}
\label{TableIsResults}
\end{table}%

\section{Radially Averaged Lattice Dispersion Relation}
\label{Section:Radial}
In the previous sections we saw that the dispersion relations on the lattice are not functions of the magnitude of the wavevector but rather on its vector decomposition in Fourier space. In some applications however, we are interested in having a radially averaged dispersion relation to match numerical data with theory. One example is the computation of the radially averaged two-point correlation function in thermal fields which requires knowledge of the radially averaged lattice dispersion relation \cite{false-vacuum}. In this section we show how to correctly calculate it.

Given a wavevector magnitude $k=|\mathbf{k}|$ we can integrate Eq.~\ref{DispersionIsotropic} over the surface of the positive octet of a sphere with radius $k$, $S:k^2=k_x^2+k_y^2+k_z^2$ and then divide by the area $A=\pi k^2/2$ to get the radially average dispersion relation. First we parametrize the sphere using

\begin{equation}
 \bm{r}(\phi,\theta)=k\cos \phi \cos \theta \bm{\hat{k}_x}+k\sin \phi \cos \theta \bm{\hat{k}_y}+k\sin \theta \bm{\hat{k}_z}
\end{equation}
with parameter space $\Omega:0\leq \phi \leq \frac{1}{2}\pi,0\leq \theta \leq \frac{1}{2}\pi$. Then the integral of Eq.~\ref{DiscreteIsotropic} over the sphere becomes

\begin{equation}
\omega^2(k)=\frac{2}{\pi k^2}\iint\limits_S \omega^2(k_x,k_y,k_z)d\sigma=\frac{2}{\pi k^2}\iint\limits_\Omega \omega^2\left(k_x(\phi,\theta),k_y(\phi,\theta),k_z(\phi,\theta)\right)||\bm{N}(\phi,\theta)||d\sigma,
\end{equation}
where $\bm{N}(\phi,\theta)=\bm{r'}_\phi(\phi,\theta)\times\bm{r'}_\theta(\phi,\theta)$ is a normal vector to the surface at the point $(\phi,\theta)$. For a sphere of radius $k$ we have $||\bm{N}(\phi,\theta)||=k^2\cos \theta$. The expression for the radially summetric dispersion relation then becomes

\begin{eqnarray}
 \omega^2(k)&=&-\frac{c_4}{a^2}-\frac{2}{\pi a^2}\int_0^{\pi/2}\cos\theta\int_{0}^{\pi/2}\big[ 2c_3\left[\cos \left[ak\cos \phi \cos \theta\right]+\cos [ak\sin \phi \cos \theta]+\cos [ak\sin \theta]\right]\nonumber \\
&& +4c_2\left[\cos \left[ak\cos \phi \cos \theta\right]\cos [ak\sin \phi \cos \theta]+\cos \left[ak\cos \phi \cos \theta\right]\cos [ak\sin \theta]+\cos [ak\sin \phi \cos \theta]\cos [ak\sin \theta]\right]\nonumber\\
&& +8c_1\cos \left[ak\cos \phi \cos \theta\right]\cos [ak\sin \phi \cos \theta] \cos [ak\sin \theta]\big]d\phi d\theta+m^2
\label{DispersionRadial}
\end{eqnarray}

\begin{figure}[htbp]
\includegraphics[scale=0.5]{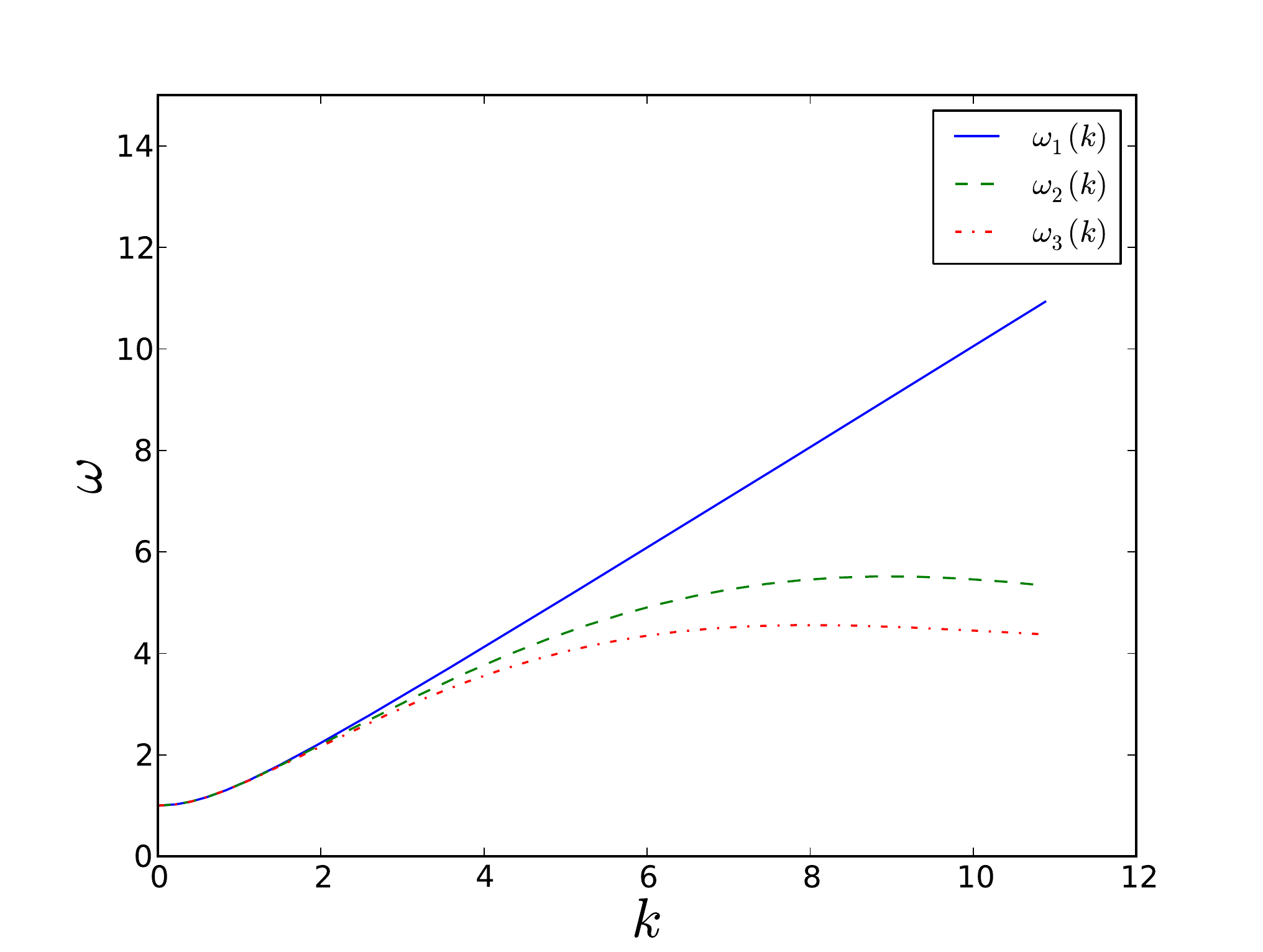}
\caption{Three dispersion relations: The continuous case with $\omega_1(k)=\sqrt{k^2+m^2}$ and two radially averaged discretizations, $\omega_2(k)$ and $\omega_3(k)$. The mode frequency $\omega$ and the wavevector magnitude $k$ are given in units of the mass $m$.}
\label{Fig:DispersionRadial}
\end{figure}

Fig.~\ref{Fig:DispersionRadial} shows the continuous dispersion relation $\omega_1(k)=\sqrt{k^2+m^2}$ and the radially symmetric dispersion relation computed from Eq.~\ref{DispersionRadial} for two discretization schemes: $\omega_2(k)$ using the anisotropic stencil with $c_1=c_2=0,c_3=1,c_4=-6$ and $\omega_3(k)$ using the isotropic stencil with $c_1=1/30,c_2=1/10,c_3=7/15,c_4=-64/15$. As expected, all three agree for low $k$ modes, but they quickly diverge from one another as $k$ gets large.

\section{Application in preheating simulations}
\label{Section:Preheating}
Even though LATTICEEASY and DEFROST use a discrete grid to perform the field evolutions, the dispersion relation employed is the continuous $\omega^2=k^2+m^2$. As it was shown in Sections~\ref{Section:Anisotropic} and \ref{Section:Isotropic}, this leads to significant discrepancies between the lattice frequency and the one predicted in the continuous limit. For this reason, we investigate how the use of the lattice dispersion relation affects previous results of preheating simulations. We reproduce the results of \cite{Felder-equilibrium} using the original LATTICEEASY with the continuous dispersion relation and a modified version in which it is correctly discretized as shown in Section~\ref{Section:Anisotropic}. We focus our attention on a chaotic inflation model with a quartic inflaton potential. The inflaton $\phi$ has a four-leg coupling to another scalar field $\chi$. The potential for this model is 

\begin{equation}
 V(\phi,\chi)=\frac{1}{4}\lambda\phi^4+\frac{1}{2}g^2\phi^2\chi^2,
\end{equation}
with equations of motion given by

\begin{equation}
 \ddot{\phi}+3\frac{\dot{a}}{a}\dot{\phi}-\frac{1}{a^2}\nabla^2\phi+(\lambda\phi^2+g^2\chi^2)\phi=0
\label{EOM:reheating1}
\end{equation}

\begin{equation}
 \ddot{\chi}+3\frac{\dot{a}}{a}\dot{\chi}-\frac{1}{a^2}\nabla^2\chi+g^2\phi^2\chi=0.
\label{EOM:reheating2}
\end{equation}
The fields are initialized as Gaussian random fields and the scale factor is evolved self-consistently by the Friedmann equations

\begin{equation}
 \left(\frac{\dot{a}}{a}\right)^2=\frac{8\pi}{3}\rho
\end{equation}

\begin{equation}
 \ddot{a}=-\frac{4\pi}{3}(\rho+3p)a
\end{equation}
where $\rho$ and $p$ refer to the energy density and pressure of the fields respectively. In LATTICEEASY, the friction terms in Eq.~\ref{EOM:reheating1} and Eq.~\ref{EOM:reheating2} are eliminated by appropriate rescalings. For details of the LATTICEEASY implementation we refer the reader to the Appendix of \cite{Felder-equilibrium}. 
\subsection{Initial Conditions}
The initial conditions are set in Fourier space and then transformed back to get the initial field values in configuration space. The simulation starts at the end of inflation and each mode is given a random phase and a gaussian distributed amplitude characterized by

\begin{equation}
 \langle|f_k|^2\rangle=\frac{1}{2\omega_k}
\label{generic_dispersion_latticeeasy}
\end{equation}
where

\begin{equation}
 \omega^2_k=k^2+m_{\textrm{eff}}^2
\label{dispersion_latticeeasy}
\end{equation}
and

\begin{equation}
 m_{\textrm{eff}}^2=\frac{\partial^2 V}{\partial f^2}=\begin{cases}
  3\lambda\langle\phi^2\rangle+g^2\langle\chi^2\rangle \\ 
  g^2\langle\phi^2\rangle 
\end{cases}					    
\end{equation}
for $\phi$ and $\chi$ respectively. Here $\langle\rangle$ denote spatial averages over the grid.

\begin{figure}[htbp]
\includegraphics[scale=0.5]{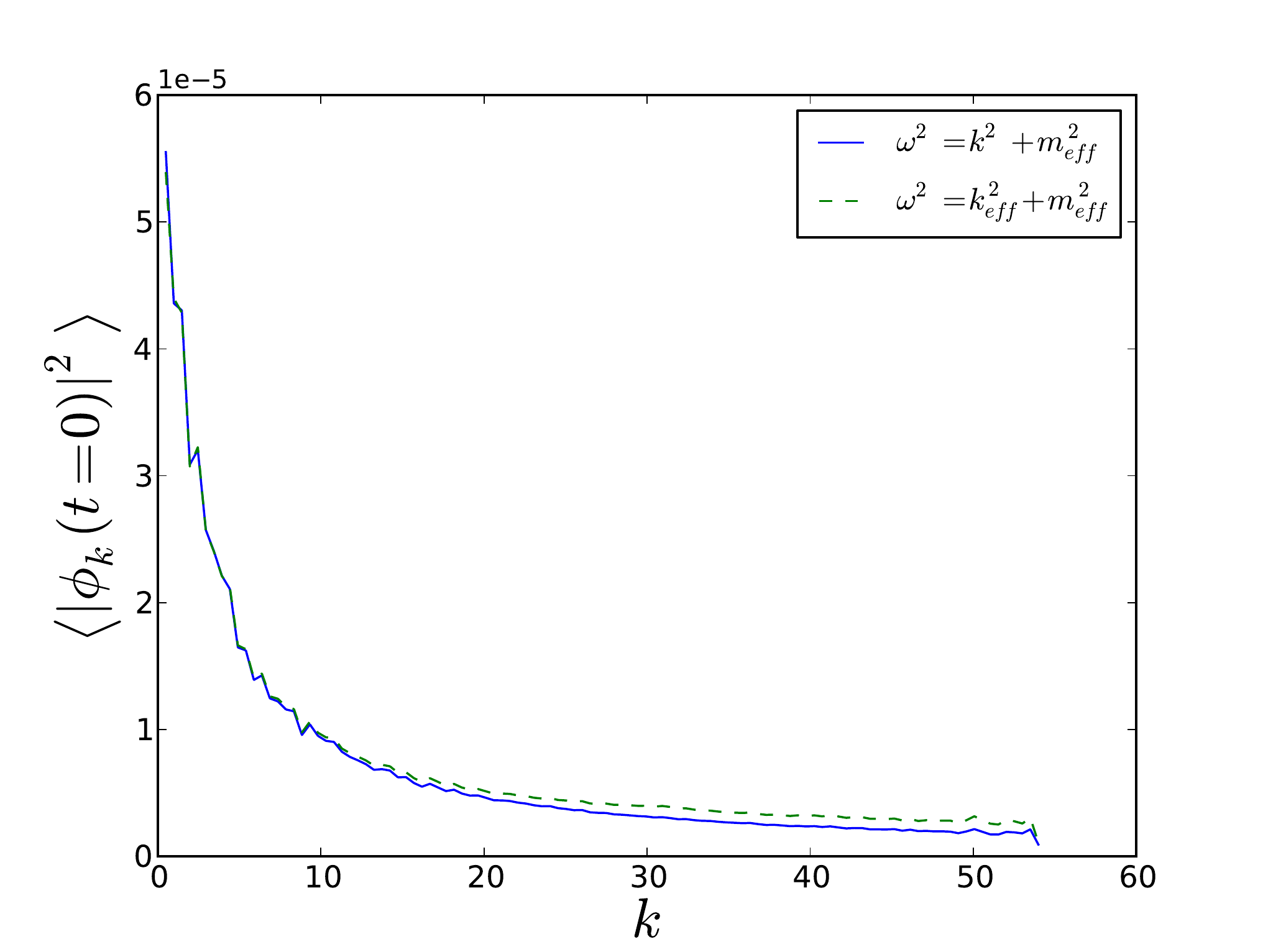}
\caption{Power spectrum of field $\phi$ for a chaotic inflation model using LATTICEEASY. The lattice dispersion relation adds more power to high $k$ modes.}
\label{Fig:initData}
\end{figure}

Since LATTICEEASY discretizes the Laplacian using the anisotropic scheme of Section\ref{Section:Anisotropic}, we compare the initial power spectrum of the fields using Eq.\ref{dispersion_latticeeasy} to what we get using $\omega^2=k_{\textrm{eff}}^2+m_{\textrm{eff}}^2$ with $k_{\textrm{eff}}^2$ given by Eq.\ref{keffAnis}. We use a box of size $128^3$ and lattice spacing $a=0.1$ and in order to be consistent with \cite{Felder-equilibrium} we use $\lambda=9\times10^{-14}$ and $g^2=200\lambda$. All quantities are measured in Planck units ($M_p=1.22\times10^{19}\textrm{GeV}$) just like in \cite{Felder-equilibrium}. The initial power spectrum using the two different dispersion relations is shown in Fig.~\ref{Fig:initData} for the field $\phi$. Modes with $k\gtrsim 10$ show a lack of power when using the continuous dispersion relation. This is not surprising if we look at Fig.~\ref{Fig:DispersionRadial}: the radially averaged $\omega_2$ is smaller than $\omega_1$ for high values of $k$ which results in larger $\langle|\phi_k|^2\rangle$ in light of Eq.~\ref{generic_dispersion_latticeeasy}. The power spectrum of $\chi$ shows a similar lack of power for high $k$ modes.

Even though the discrepancy in the initial conditions is evident, we must examine whether it has any effect on the process of preheating. For this reason, we evolve the fields until the system reaches thermalization and compare the occupation numbers of the fields for the continuous and discrete dispersion relations.
\subsection{Occupation Number Density}
\begin{figure}[htbp]
\includegraphics[scale=0.5]{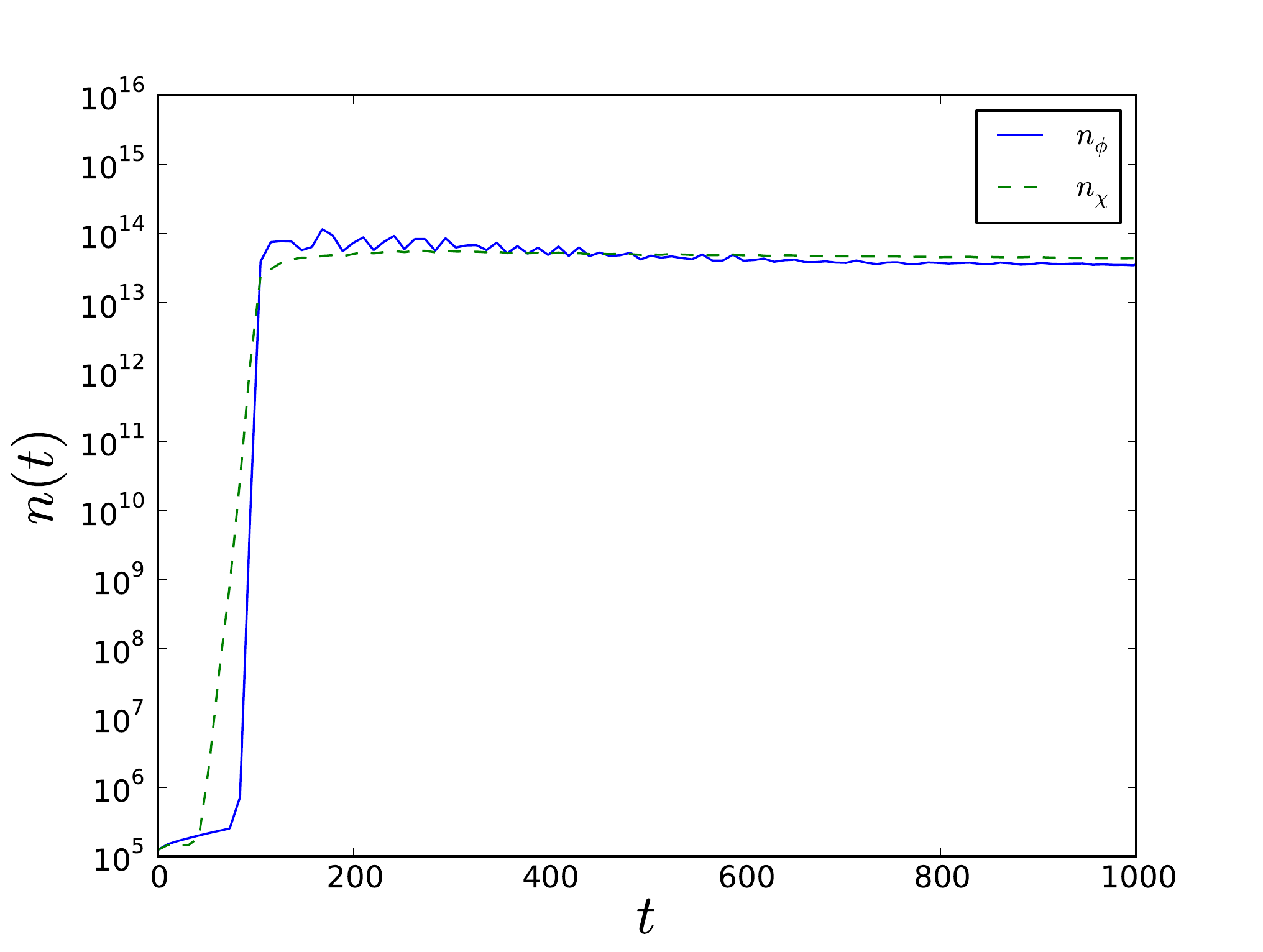}
\caption{Comoving number of particles as a function of time during reheating for fields $\phi$ and $\chi$. Parametric resonance in the fields induces exponential growth at early times which then levels off as thermalization is approached.}
\label{Fig:total_occup_number}
\end{figure}

\begin{figure}[htbp]
\includegraphics[scale=0.5]{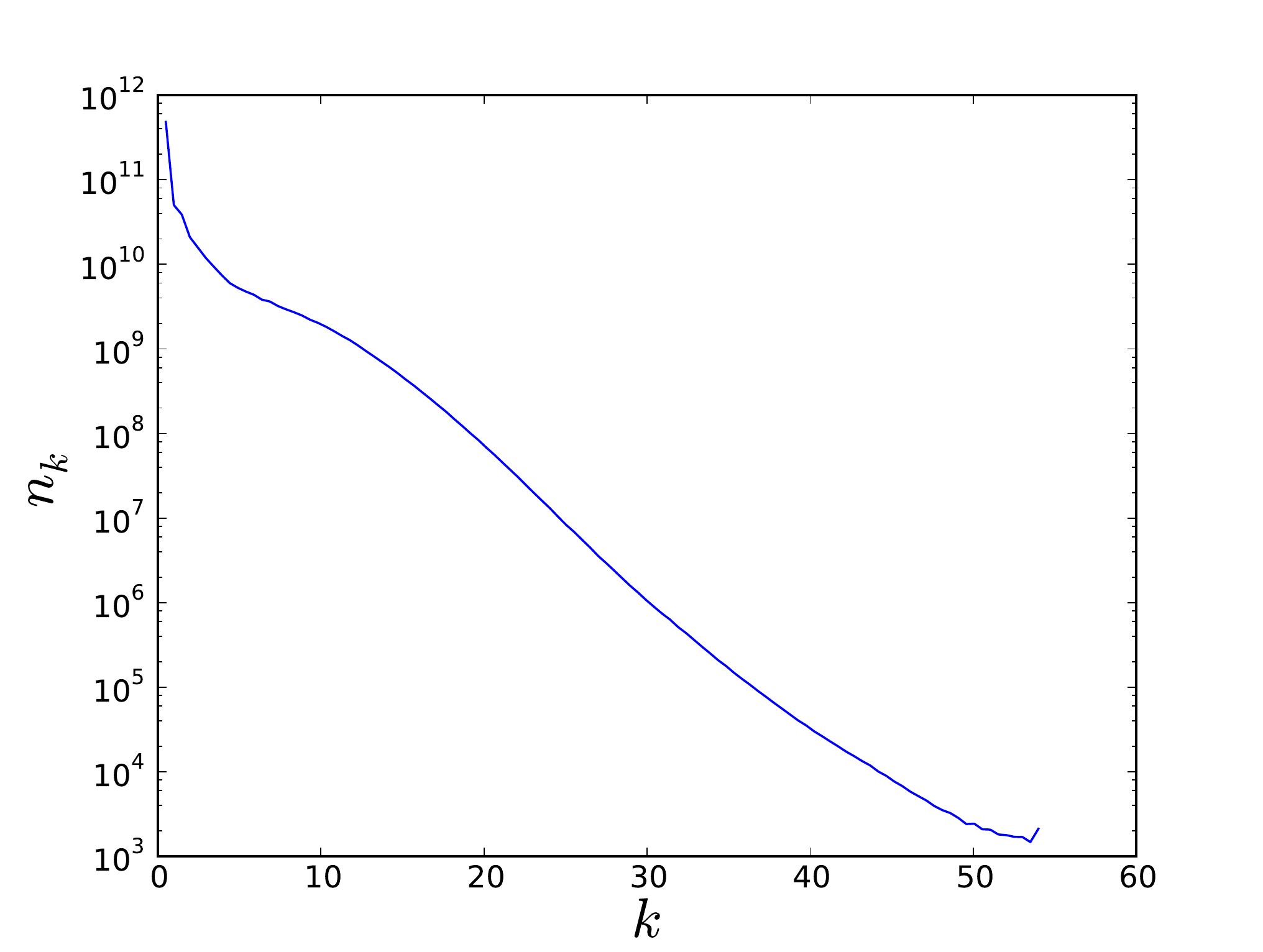}
\caption{Particle number density for the field $\phi$ at $t=1000$.}
\label{Fig:occupation_number_density_full}
\end{figure}

The most important variable to calculate during preheating is the comoving number of particles in each field defined by

\begin{equation}
 n_f(t)\equiv\frac{1}{(2\pi)^3}\int d^3k n_k(t),
\end{equation}
where $n_k$ is the comoving occupation number density of particles

\begin{equation}
 n_k(t)\equiv \frac{1}{2\omega_k}|\dot{f}_k|^2+\frac{\omega_k}{2}|f_k|^2.
\label{number_density}
\end{equation}
During preheating, the number of particles in each field undergoes exponential growth induced by parametric resonance. A convenient way to label the end of preheating is by looking at the time when $n(t)$ levels off for each field. Fig.~\ref{Fig:total_occup_number} shows the evolution of $n(t)$ for both fields $\phi$ and $\chi$ and matches Fig.~13 of \cite{Felder-equilibrium}. In order to check how the lattice dispersion relation affects preheating, we focus on a time that both fields are long past the exponential growth regime, $t=1000$ and compare the particle number density. Fig~\ref{Fig:occupation_number_density_full} shows the number density for the field $\phi$ at time $t=1000$.

\begin{figure}[htbp]
\includegraphics[width=0.49\linewidth]{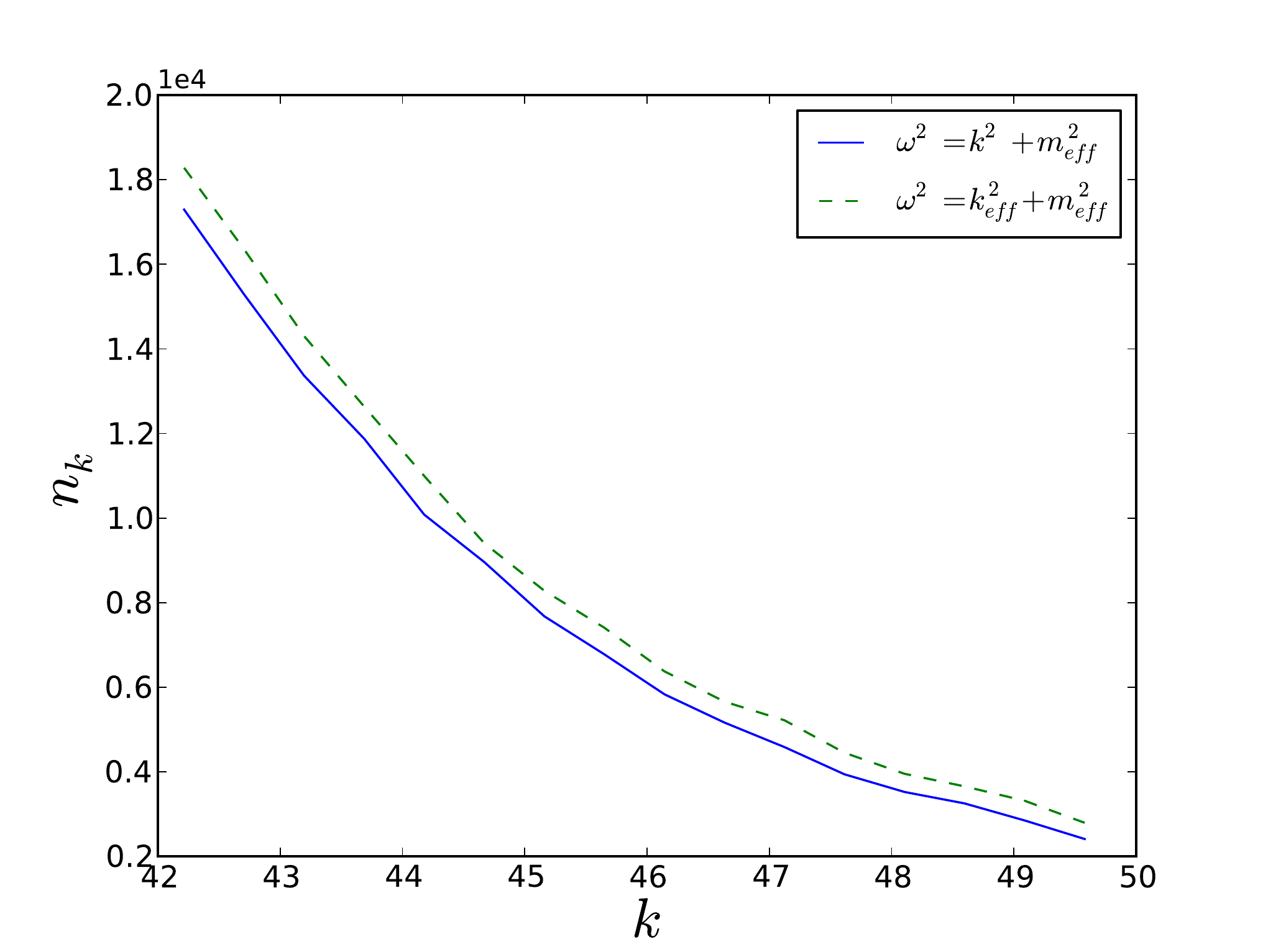}
\includegraphics[width=0.49\linewidth]{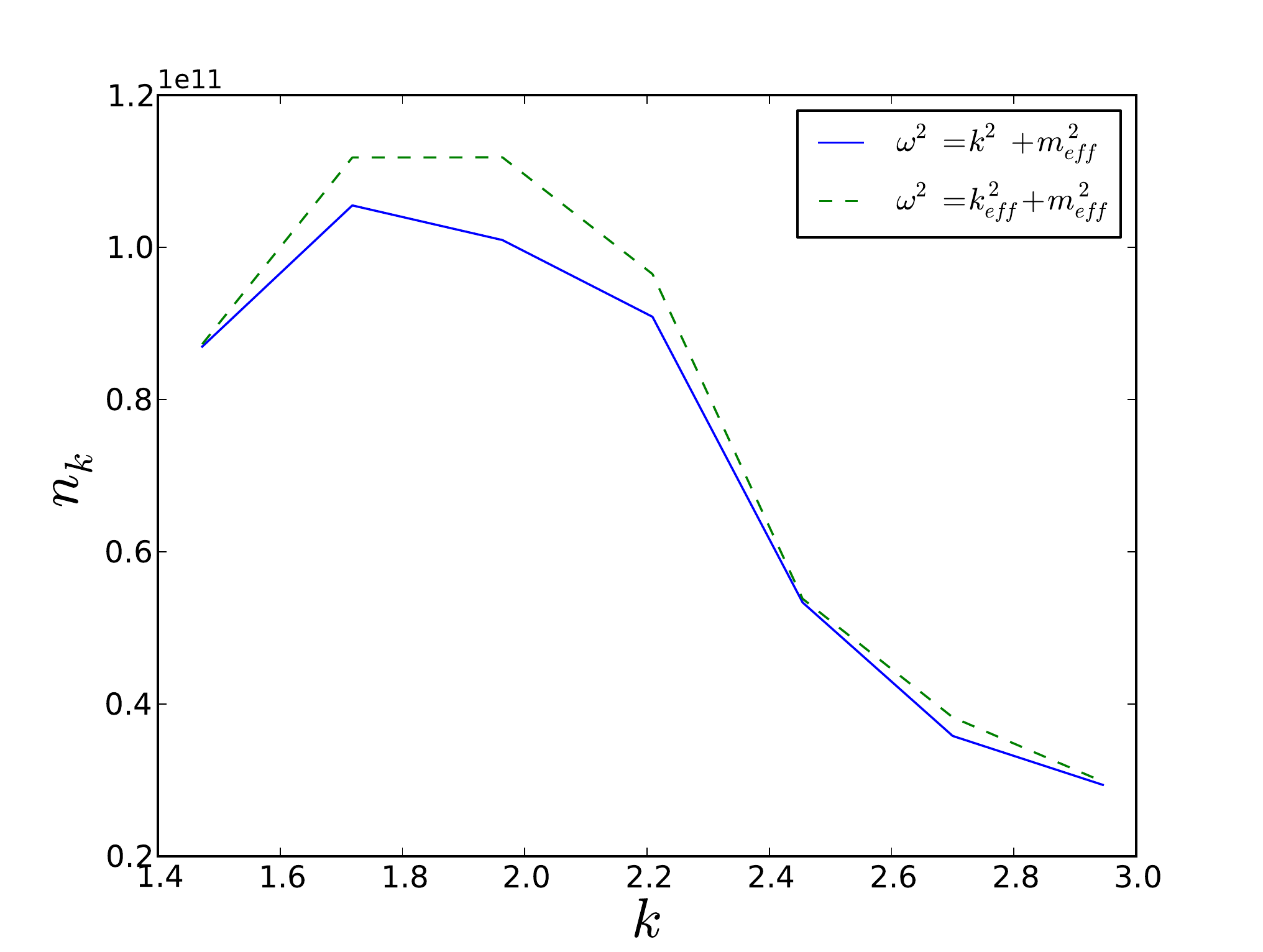}
\caption{Lattice dispersion relation effects on the particle number density for the field $\phi$ at $t=1000$, shown for small and large values of the wavevector magnitude $k$.}
\label{Fig:occupation_number_density_split}
\end{figure}

We have performed the run first using the original dispersion relation for both the initial conditions and the calculation of the particle number density in Eq.~\ref{number_density} and then using the lattice dispersion relation $\omega^2=k_{\textrm{eff}}^2+m_{\textrm{eff}}^2$ with $k_{\textrm{eff}}^2$ given by Eq.\ref{keffAnis}. The comparison of the two methods is shown in Fig.~\ref{Fig:occupation_number_density_split} for ranges of $k$ in both the low and high end of the spectrum which are most affected by the different dispersion relations. The left part of the figure shows that the initial lack of power in high $k$ modes arising from the original dispersion relation has persisted even after the end of preheating, giving a consistently wrong number density of the order of 10\%. The most interesting and unexpected result however is for low $k$: For the range of $1.5 \lesssim k \lesssim 3.0$, both dispersion relations give almost identical mode frequencies, leading to the same initial conditions and definitions for Eq.~\ref{number_density}. Yet, even in this case, there are features which differ by as much as 10\%. This indicates that the lack of power in the initial conditions for high $k$ modes affects the process of parametric resonance, leading to an incorrect picture of particle number density even long after preheating is over. We therefore conclude that the correct adoption of the lattice dispersion relation in preheating simulations is not only desired from a consistency point of view, but required to capture all the available features of the underlying theory.

\section{Discussion and conclusions}
\label{Section:conclusions}
We have calculated the correct lattice dispersion relation for three dimensional simulations employing finite difference isotropic and anisotropic discretization methods. We have shown that, on the lattice, the frequency of a mode should not be treated as a function of its wavevector magnitude but of its vector decomposition. Moreover, we showed how to compute a radially averaged lattice dispersion relation which is important in applications where the numerically calculated two-point correlation function is to be matched to the theoretically predicted spectrum. 

Finally, we have shown that the incorrect use of the continuous dispersion relation in numerical simulations of preheating after inflation leads to a lack of power in the ultraviolet spectrum of the initial conditions which propagates to later times through the evolution of the fields. Consequently, we notice a discrepancy in previous calculations of particle number density during thermalization which is accentuated by the incorrect definition of the number density operator.

Even though we have only explored the effects of lattice dispersion relations in preheating simulations, we note that modified versions of LATTICEEASY are employed in a variety of other studies including bubble nucleation \cite{easther-bubbles}, gravitational wave production \cite{gravity_waves} and generation of non-gaussianities \cite{non-gaussianity}. Extra care should be used also in these studies to avoid inaccuracies induced by using the continuous dispersion relation on the lattice, particularly in (but not limited to) the ultraviolet spectrum. A more detailed study of these effects is left for future work.

\section{Acknowledgements}
The author would like to thank Marcelo Gleiser, Noah Graham and Peter Cuadrilla for critical feedback. NS is a Gordon F. Hull Fellow at Dartmouth College.


\begin{thebibliography}{99}
\bibitem{smit} J. Smit, \emph{Introduction to Quantum Fields on a Lattice}, Cambridge University Press; Cambridge, UK (2001).
\bibitem{numerical_qft} M. Williams, Comput. Phys. Comm. {\bf 180}, 1847 (2009).
\bibitem{black_hole_sim} W. Tichy, B. Bruegmann, M. Campanelli, P. Diener, Phys. Rev. D{\bf 67}, 064008 (2003).
\bibitem{basset} B. A. Bassett, S. Tsujikawa and D. Wands, Rev. Mod. Phys. \textbf{78}, 537 (2006).
\bibitem{Felder-equilibrium} G. Felder and L. Kofman, Phys. Rev. D{\bf 63}, 103503 (2001).
\bibitem{preheating} G. Felder, J. Garcia-Bellido, P. B. Greene, L. Kofman, A. Linde, I. Tkachev, Phys. Rev. Lett. {\bf 87}, 011601 (2001), Gary N. Felder, Lev Kofman,  Phys. Rev. D{\bf 75}, 043518 (2007).
\bibitem{oscillons1} I. L. Bogolubsky and V. G. Makhankov, JETP Lett. {\bf 24}, 12 (1976) [Pis'ma Zh. Eksp. Teor. Fiz. {\bf 24}, 15 (1976)].
\bibitem{oscillons2} M. Gleiser, Phys. Rev. D{\bf 49}, 2978 (1994).
\bibitem{oscillons3} E. J. Copeland, M. Gleiser and H.-R. M\"uller, Phys. Rev. D{\bf 52}, 1920 (1995).
\bibitem{oscillons_cosm1} N. Graham and N. Stamatopoulos, Phys. Lett. B{\bf 639}, 541 (2006).
\bibitem{oscillons_cosm2} E. Farhi, N. Graham, A. Guth, N. Iqbal, R. Rosales and N. Stamatopoulos, Phys. Rev. D {\bf 77}, 085019 (2008).
\bibitem{oscillons_cosm3} M. Gleiser, N. Graham, N. Stamatopoulos, Phys. Rev. D{\bf 82}, 043517, (2010).
\bibitem{amin_oscillons} Mustafa A. Amin, Richard Easther, Hal Finkel, JCAP {\bf 1012}, 001 (2010).
\bibitem{marcelo_oscillons} Marcelo Gleiser, Noah Graham, Nikitas Stamatopoulos, Phys. Rev. D{\bf 83}, 096010 (2011).
\bibitem{amin_oscillons2} Mustafa A. Amin, Richard Easther, Hal Finkel, Raphael Flauger, Mark P. Hertzberg, [arXiv:1106.3335].
\bibitem{joel_higgs} M. Gleiser, J. Thorarinson, Phys. Rev. D{\bf 79}, 025016 (2009).
\bibitem{noah_electroweak} N. Graham, Phys. Rev. Lett. {\bf 98}, 101801 (2007), [Erratum-ibid. {\bf 98}, 189904 (2007)]; Phys. Rev. D{\bf
76} (2007) 085017.
\bibitem{latticeeasy} G. Felder and I. Tkachev, Comput. Phys. Commun. {\bf 178}, 929 (2008).
\bibitem{defrost} Andrei V. Frolov, JCAP {\bf 0811}, 009 (2008).
\bibitem{pspectre} Richard Easther, Hal Finkel, Nathaniel Roth, [arXiv:1005.1921].
\bibitem{false-vacuum} Marcelo Gleiser, Barrett Rogers, Joel Thorarinson, Phys. Rev. D{\bf 77}, 023513 (2008).
\bibitem{hindmarsh} Mark Hindmarsh, Petja Salmi, Phys. Rev. D{\bf77}, 105025 (2008); Phys. Rev. D{\bf 74}, 105005 (2006). 
\bibitem{shaposhnikov} K. Kajantie, M. Laine, K. Rummukainen and M. Shaposhnikov, Phys. Rev. Lett. {\bf 77}, 2887 (1996); Nucl. Phys. B{\bf 493}, 413 (1997).
\bibitem{isotropic} M. Patra and M. Karttunen, Num. Meth. for PDEs {\bf 22}, 936, (2005).
\bibitem{easther-bubbles} R. Easther, J. T. Giblin Jr, L. Hui, E. A. Lim,  Phys. Rev. D{\bf 80}, 123519 (2009).
\bibitem{gravity_waves} J. Dufaux, D. G. Figueroa, J. Garcia-Bellido, Phys. Rev. D{\bf 82}, 083518 (2010), L. R. Price, X. Siemens, Phys. Rev. D{\bf 78}, 063541 (2008), R. Easther, J. T. Giblin Jr, E. A. Lim, Phys. Rev. D{\bf 77}, 103519 (2008).
\bibitem{non-gaussianity} A. Chambers, S. Nurmi, A. Rajantie, JCAP {\bf 1001}, 012 (2010).
\end{thebibliography}
\end{document}